\pdfoutput=1

\documentclass[aps,pre,10pt,twoside,twocolumn,floatfix]{revtex4-1}

\usepackage{amsmath,amssymb,bm,mathptmx,graphicx}

\usepackage[colorlinks,linkcolor=blue,citecolor=blue,urlcolor=blue]{hyperref}
\usepackage[OT2,T1]{fontenc}
\usepackage[russian,english]{babel}
\DeclareRobustCommand{\cyril}[1]%
{\begingroup\fontfamily{erewhon-TLF}\foreignlanguage{russian}{#1}\endgroup}

\def\leq{\ensuremath{\leqslant}}
\def\geq{\ensuremath{\geqslant}}
\def\abs#1{\ensuremath{\lvert{#1}\rvert}}

\begin{document}

\title{Approximate probabilistic cellular automata for the dynamics of single-species populations \\ under discrete logisticlike growth with and without weak Allee effects}

\author{J. Ricardo G. Mendon\c{c}a}
\email[Email: ]{jricardo@usp.br}
\affiliation{Escola de Artes, Ci\^{e}ncias e Humanidades, Universidade de S\~{a}o Paulo \\ Rua Arlindo Bettio 1000, Ermelino Matarazzo, 03828-000 S\~{a}o Paulo, SP, Brazil}

\author{Yeva Gevorgyan}
\email[Email: ]{yeva@ime.usp.br}
\affiliation{Departamento de Matem\'{a}tica Aplicada, Instituto de Matem\'{a}tica e Estat\'{\i}stica, Universidade de S\~{a}o Paulo \\ Rua do Mat\~{a}o 1010, Cidade Universit\'{a}ria, 05508-090 S\~{a}o Paulo, SP, Brazil}

\begin{abstract}
We investigate one-dimensional elementary probabilistic cellular automata (PCA) whose dynamics in first-order mean-field approximation yields discrete logisticlike growth models for a single-species unstructured population with nonoverlapping generations. Beginning with a general six-parameter model, we find constraints on the transition probabilities of the PCA that guarantee that the ensuing approximations make sense in terms of population dynamics and classify the valid combinations thereof. Several possible models display a negative cubic term that can be interpreted as a weak Allee factor. We also investigate the conditions under which a one-parameter PCA derived from the more general six-parameter model can generate valid population growth dynamics. Numerical simulations illustrate the behavior of some of the PCA found. \\

{\noindent}\href{https://doi.org/10.1103/PhysRevE.95.052131}{DOI: 10.1103/PhysRevE.95.052131} 
\end{abstract}

\pacs{87.23.Cc, 05.40.-a, 02.50.-r}

\keywords{population dynamics, Allee effect, logistic map, cubic map, cellular automata, mean-field approximation}

\maketitle


\section{\label{intro}Introduction}

Cellular automata (CA) are discrete-space, discrete-time deterministic dynamical systems that map symbols from a finite set of symbols into the same set of symbols according to some well-defined set of rules. Usually, the discrete dynamical cells that characterize and compose the CA update their states simultaneously, the underlying lattice of cells is regular---like, e.\,g., an array of equally spaced points or a square or hexagonal lattice---and the dynamics of each cell depends only on the states of other cells over a finite neighborhood. The idea of CA dates back at least to the end of the 1940s, when they were conceived as model systems for simple self-reproducing, self-repairing organisms and, by extension, logical elements and memory storage devices
\cite{burks}. For broad introductions to the subject see
\cite{wolfram,toffoli,chopard,boccara,adamatzky}.

A CA with rules depending on a random variable becomes a probabilistic CA (PCA). PCA were introduced mainly by the Russian school of stochastic processes in the decade of 1960--1970 in relation with the positive probabilities conjecture---a conjecture that is deeply rooted in the theory of Markov processes and has a counterpart in the well-known statistical physics lore that one-dimensional systems do not display phase transitions at finite ($T>0$) temperature---, but also as model systems for noisy ``neurons'' and voting systems
\cite{dobrushin,petrovskaya,classification,stavskaya,toom,gkl,russkiye} (see also \cite{ledoussal,lebowitz}). Besides serving as model systems for the analysis of computation, both applied and theoretical, digital or biological, CA and PCA have also been playing a significant role in the modeling of biological and ecological complex systems, notably spatial processes and the interplay between dispersion and competition in the determination of structure and scale of ecosystems \cite{hogeweg,silvertown,camodel,durrett,britaldo,green,abmbook}.

In this paper we investigate how microscopic PCA models of a single-species unstructured population with nonoverlapping generations average in first-order, single-cell mean-field approximation to well-known models widely employed in population dynamics, namely, the logistic map and a variant cubic map that describes the dynamics of a population under weak Allee effects. The mean-field equations provide a connection between the microscopic stochastic dynamics of the PCA with the deterministic dynamics of the ensuing discrete-time maps, that otherwise are typically derived using \textit{ad hoc} (phenomenological, at best) arguments.

The probabilistic nature of the PCA together with the functional form of the logistic and cubic maps that we want to recover from the mean-field approximations impose constraints that the microscopic transition probabilities have to observe to make the models sensible from the point of view of population dynamics. Otherwise, the construction of the microscopic PCA models with the desired population dynamics properties from the elementary transitions available is straightforward. We believe that the fact that both the logistic map and a cubic map that incorporates the description of weak Allee effects can be obtained from a single, relatively simple microscopic framework is worth notice. Approaches closely related to ours have been employed in \cite{cellscale,mcelwain}; their interacting particle systems, however, evolve in continuous time and are not homogeneous in space. A somewhat related approach has also appeared in the sociophysics literature \cite{bagrecht}.

This paper is structured as follows. In Sec.~\ref{sec:log} we review the logistic growth model in discrete time for the dynamics of a single-species unstructured population with nonoverlapping generations and one of its possible extensions to include Allee effects. Section~\ref{sec:pca} presents the PCA formalism and the mean-field approximation to their dynamics. In this section we obtain the first-order, single-cell mean-field approximation to the most general one-dimensional left-right symmetric PCA that is central to our analysis. In Sec.~\ref{sec:pcalog} we derive the constraints on the transition probabilities of the PCA under which its six-parameter mean-field approximation yields the logistic map often encountered in population dynamics as well as an extension thereof containing a cubic term that models weak Allee effects. There we also obtain the solution sets in the six-parameter (actually, five-parameter) space that encompass the models of interest, present simulations of some of the PCA models obtained, and discuss their main features. In Sec.~\ref{sec:pq} we examine the models that survive the constraints when we reduce the number of parameters of the PCA from six (actually, five) to a single one. We found several, some of which can be understood as probabilistic mixtures of elementary CA. Finally, in Sec.~\ref{sec:summary} we assess the results obtained and indicate possible directions for further research. An Appendix details some calculations in the mean-field approximation for some of the single-parameter PCA found.


\section{\label{sec:log}Discrete logisticlike growth models}

\subsection{\label{discrete}The discrete logistic growth model}

The dynamics of a single-species population subject to limiting resources can be described by Verhulst's logistic growth model \cite{verhulst}
\begin{equation}
\label{eq:log}
\frac{d{x}_{t}}{dt} = x_{t}g(x_{t}) = rx_{t}\Big(1-\frac{x_{t}}{K}\Big),
\end{equation}
where $x_{t} \geq 0$ represents the size of the population, $r>0$ is the maximum potential rate of reproduction of the individuals in the population, and $K>0$ is the carrying capacity, defined as the maximum population viable under the given ecological conditions \cite{ekeshet,brauer}. The function $g(x_{t})$ represents the instrinsic growth rate per capita of the population, and the form of the right-hand side of (\ref{eq:log}) guarantees that when $x_{t}=0$ there is no spontaneous generation of living organisms. The intrinsic growth rate $g(x_{t})=r(1-x_{t}/K)$ decreases to zero as the population grows large, and arose as an improvement over the simple model of constant growth rate to make it more realistic. The basic model (\ref{eq:log}) has been generalized in several different ways to account for particular conditions and factors affecting specific populations \cite{tsoularis}.

Discrete-time models are particularly suited to describe the dynamics of populations with nonoverlapping generations, in which the population growth takes place at discrete intervals of time \cite{may1974,hassell}. This is typically the case for populations of annual plants and insects. The discrete-time version of (\ref{eq:log}), a.\,k.\,a.\ the logistic map, is given by
\begin{equation}
\label{eq:map}
x_{t+1} = rx_{t}\left(1-\frac{x_{t}}{K}\right).
\end{equation}
Equation (\ref{eq:map}) is sometimes presented in the form $x_{t+1}-x_{t}=rx_{t}(1-x_{t}/K)$, which is perhaps biologically more revealing; the two forms, however, are equivalent by the redefinitions $r \leftrightarrow (1+r)$ and $K \leftrightarrow (1+1/r)K$. The logistic map (\ref{eq:map}) displays much more complex a behavior than (\ref{eq:log}), having become the archetypal example of a chaotic dynamical system
\cite{may1974,hassell,may1976a,may1976b,iterated,devaney}.

In this work we are interested in the logistic map both in its basic form (\ref{eq:map}) and with an extra term included to take into account Allee effects, which we introduce in the next subsection.


\subsection{\label{sec:allee}Allee effects}

The Allee effect, first discussed in the 1930's, describes a positive correlation between the rate of growth and the density of a population, namely, that at low population densities reproduction and survival of individuals may decline \cite{allee,dennis,lande,stephens}. The effect saturates or disappears as populations grow larger. The Allee effect challenges the classical tenet of population dynamics according to which individual fitness is higher at low densities because of lower intraspecific competition. Allee effects can be related with components of individual fitness (the component Allee effect) or with overall mean individual fitness (the demographic Allee effect), which is what in general can be measured in the field. Empirical evidence suggests that Allee effects are caused mainly by mate limitation, debilitated cooperative defense, unsubstantial predator satiation, lack of cooperative feeding, dispersal, and habitat alteration \cite{gascoigne,kramer}. Recently, it has been argued that tumour growth displays many features of population dynamics, including Allee effects \cite{korolev}.

\begin{figure}
\centering
\includegraphics[viewport=80 40 450 415, scale=0.30, clip]{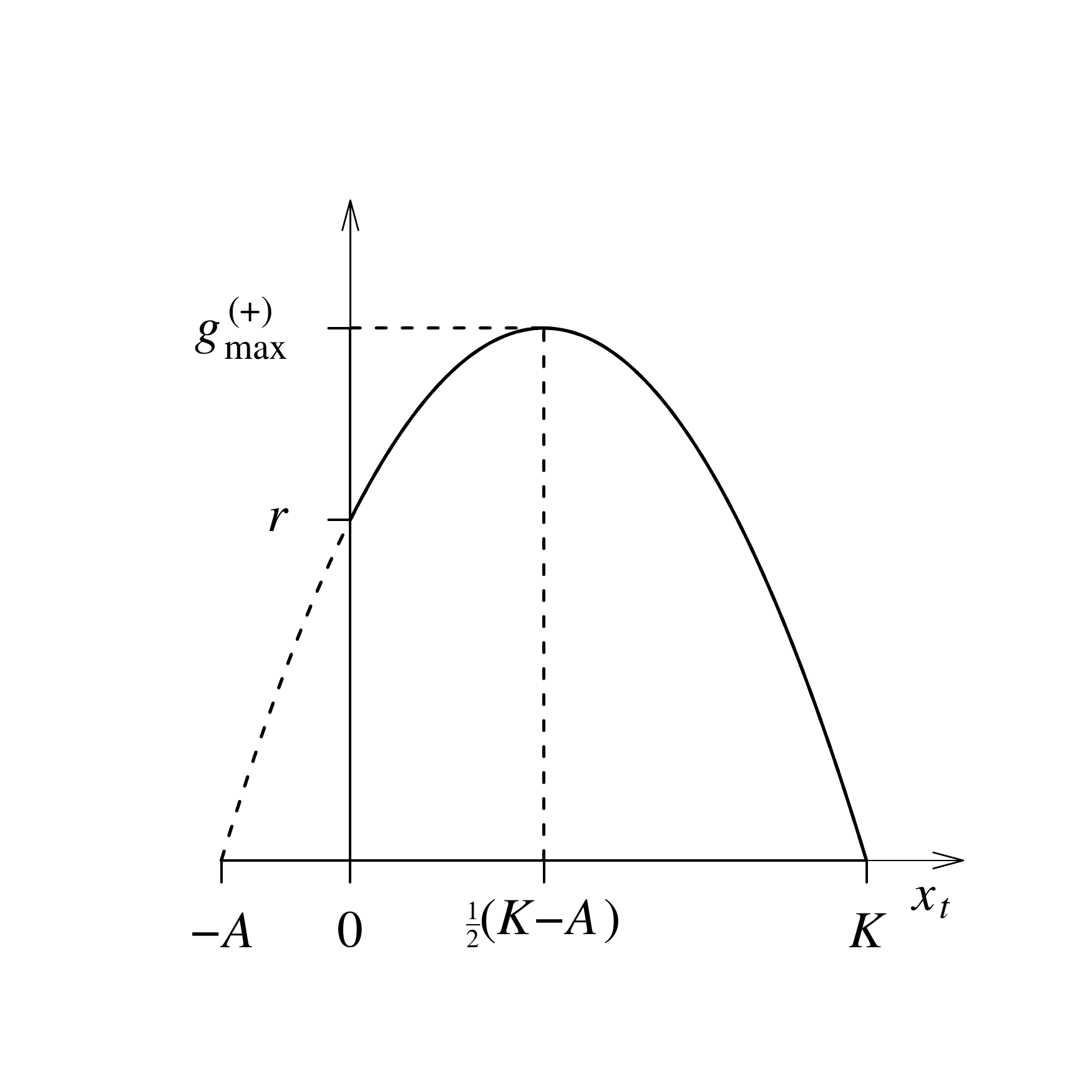} \hfill
\includegraphics[viewport=30 40 450 415, scale=0.30, clip]{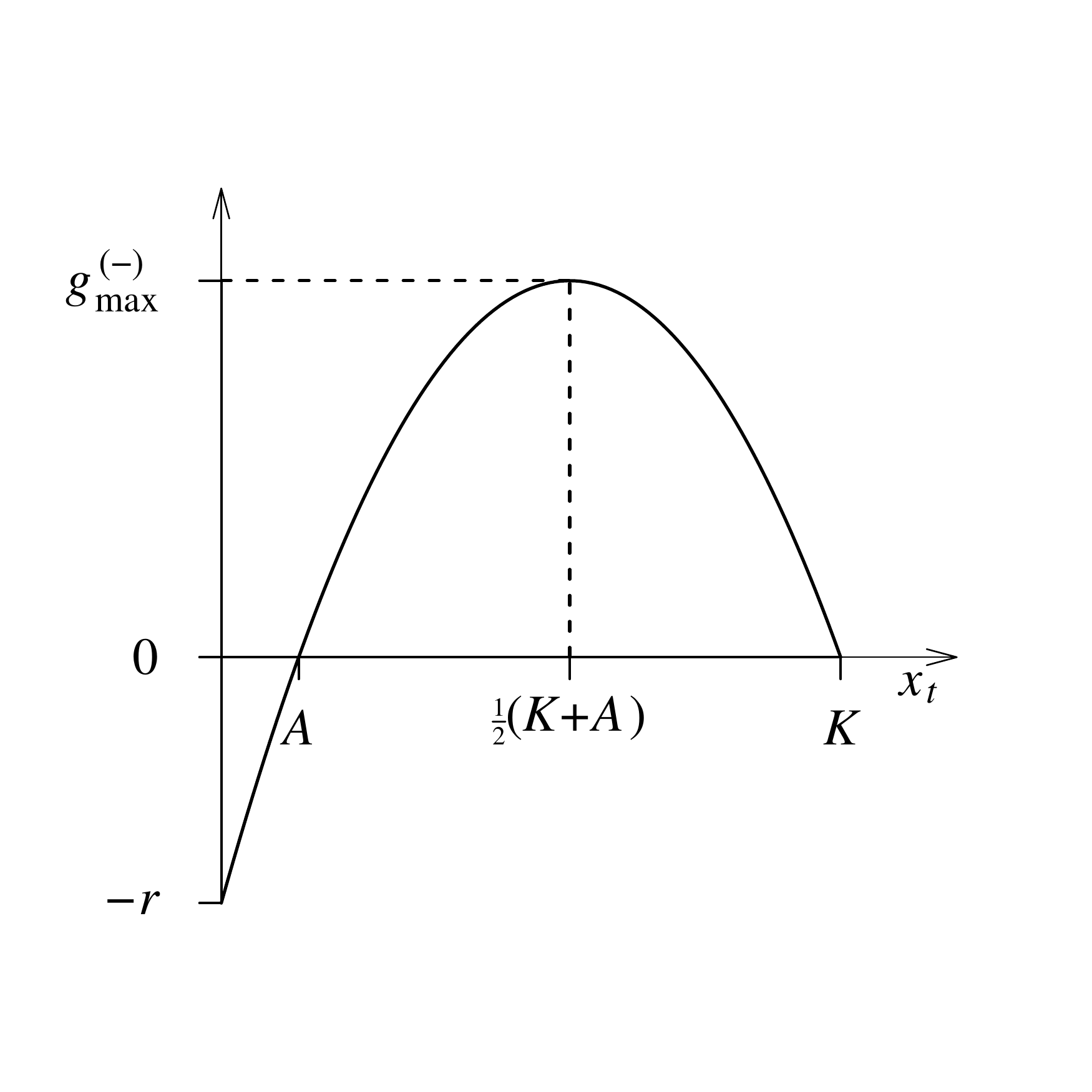}
\caption{\label{fig:ggg}Intrinsic population growth rate functions $g^{(+)}(x_{t})$ (left) and $g^{(-)}(x_{t})$ (right) given by (\ref{eq:cubic}). These functions provide models for the weak [$g^{(+)}(x_{t})$] and strong [$g^{(-)}(x_{t})$] Allee effects.}
\end{figure}

It is possible to model the demographic Allee effect in several different ways, following different biological rationale \cite{boukal}. Mathematically, the Allee effect requires that $g(x_{t})$ have a maximum at intermediate densities and decay at low densities. It is also desirable that $g(x_{t})$ decays at high densities to display the logistic effect. The simplest way to obtain this combined behavior is by multiplying the righ-hand side of (\ref{eq:log}) or (\ref{eq:map}) by $(x_{t}/A \pm 1)$, where $A$ represents a critical population threshold \cite{kareiva,hastings,gruntfest,courchamp,bascompte}. The resulting growth rate function becomes
\begin{equation}
\label{eq:cubic}
g^{(\pm)}(x_{t}) = r\Big(1-\frac{x_{t}}{K}\Big)\Big(\frac{x_{t}}{A}\pm 1\Big),
\end{equation}
with $r$, $K$, and $A<K$ positive constants. Functions (\ref{eq:cubic}) are concave down parabolas with roots $g^{(\pm)}(K)=g^{(\pm)}(\mp A)=0$ and maximum at $x_{\mathrm{max}}^{(\pm)}=\frac{1}{2}(K \mp A)$, where $g^{(\pm)}(x_{\mathrm{max}}^{(\pm)})=$ $r(K \pm A)^{2}/4KA >0$. These functions are depicted in Fig.~\ref{fig:ggg}.

Functions $g^{(\pm)}(x_{t})$ model what are known as the weak (or non-critical) Allee effect [$g^{(+)}(x_{t})$], according to which at small population sizes the growth rate decreases but remains positive, and the strong (or critical) Allee effect [$g^{(-)}(x_{t})$], when the growth rate may become negative at small population sizes and lead the population to extinction. In this work we only consider models for the weak Allee effect, since in the PCA scenario the constant term of the intrinsic population growth rate function becomes a sum of nonnegative probabilities, therefore enforcing the $g^{(+)}(x_{t})$ model (see Sec.~\ref{sec:pcalog}).


\section{\label{sec:pca}Probabilistic cellular automata}

A one-dimensional, two-state PCA \cite{wolfram,toffoli,chopard,boccara,adamatzky,russkiye,ledoussal,lebowitz} is defined by an array of cells arranged in a one-dimensional lattice $\Lambda=\{1, 2, \dots, L\} \subset \mathbb{Z}$ of total length $L$, usually under periodic boundary condition ($L+1 \equiv 1$ and $0 \equiv L$), with each cell in one of two possible states, say, $x_{i}=0$ or $1$, $i=1,\dots,L$. The state of the PCA at instant $t=0$, $1, \dots$ is given by $\bm{x}^{t}=(x_{1}^{t},\,x_{2}^{t},\,\dots,\,x_{L}^{t}) \in \Omega = \{0,1\}^{\Lambda}$.

The probability distribution $P_{t}(\bm{x})$ of observing a particular state $\bm{x}$ of the PCA at instant $t$ given an initial distribution $P_{0}(\bm{x})$ is given by
\begin{equation}
\label{eq:markov}
P_{t+1}(\bm{x}') =
\sum_{\bm{x}\, \in\, \Omega}\Phi(\bm{x}'\,|\,\bm{x})P_{t}(\bm{x}),
\end{equation}
where $0 \leq \Phi(\bm{x}'\,|\,\bm{x}) \leq 1$ is the conditional probability for the transition $\bm{x} \to \bm{x}'$ to occur in one time step. The rules that map the state of $x_{i}^{t}$ into the new state $x_{i}^{t+1}$ depend only on a finite neighborhood of $x_{i}^{t}$. In this work the neighborhood is given by the $i$-th cell itself together with its two nearest-neighbor cells $i \pm 1$, and since the cells of the PCA are updated simultaneously and independently we have that
\begin{equation}
\label{eq:probs}
\Phi(\bm{x}'\,|\,\bm{x}) =
\prod_{i=1}^{L}\phi(x_{i}'\,|\,x_{i-1},\,x_{i},\,x_{i+1}).
\end{equation}

From Eqs.~(\ref{eq:markov}) and (\ref{eq:probs}), it is easy to show that the dynamics of the (marginal) probability distribution $P_{t+1}(x)$ of observing a cell---any cell, since the PCA is spatially homogeneous---in state $x$ at instant $t$ (equivalently, the instantaneous density of cells in state $x$ in the PCA) obeys
\begin{equation}
\label{eq:p1}
P_{t+1}(x_{i}') =
\sum_{x_{i-1},\,x_{i},\,x_{i+1}}\phi(x_{i}'\,|\,x_{i-1},\,x_{i},\,x_{i+1})
P_{t}(x_{i-1},\,x_{i},\,x_{i+1}).
\end{equation}
From this equation we see that the determination of $P_{t}(x_{i})$ depends on the knowledge of the probabilities $P_{t}(x_{i-1},\,x_{i},\,x_{i+1})$, which in turn depend on $P_{t}(x_{i-2},\,x_{i-1},\,x_{i},\,x_{i+1},\,x_{i+2})$ and so on. The simplest approach to break the hierarchy of coupled equations implied by (\ref{eq:p1}) and get a closed set of equations is to approximate
\begin{equation}
\label{eq:mf}
P_{t}(x_{i-1},\,x_{i},\,x_{i+1}) \approx
P_{t}(x_{i-1})\,P_{t}(x_{i})\,P_{t}(x_{i+1}).
\end{equation}
This is the first-order, single-cell mean-field approximation, which assumes probabilistic independence between the cells. It is possible to resort to higher order approximations involving pairs, triplets, etc.\ of cells to obtain increasingly better approximate descriptions of the dynamics of the PCA \cite{local,cluster,dickman,haye,arXiv,nazim}, but we will limit ourselves to the simple approximation (\ref{eq:mf}). As we will see, already at this level of approximation we obtain interesting models of single-species population dynamics.

A logical choice for most models of natural phenomena are left-right symmetric rules, in which case $\phi(x_{i}'\,|\,x_{i-1},x_{i},x_{i+1}) = \phi(x_{i}'\,|\,x_{i+1},x_{i},x_{i-1})$ and we are left with six transition probabilities to specify. We will denote these transition probabilities by $a$, $b$, \ldots, $f \in [0,1]$ according to the following convention:
\begin{subequations}
\label{eq:abcdef}
\begin{align}
\label{eq:a}
\phi(1\,|\,000) &= a, \\
\label{eq:b}
\phi(1\,|\,001) &= \phi(1\,|\,100) = b, \\
\label{eq:c}
\phi(1\,|\,010) &= c, \\
\label{eq:d}
\phi(1\,|\,011) &= \phi(1\,|\,110) = d, \\
\label{eq:e}
\phi(1\,|\,101) &= e, \\
\label{eq:f}
\phi(1\,|\,111) &= f.
\end{align}
\end{subequations}
Table~\ref{tab:abcdef} displays the rule table for PCA (\ref{eq:a})--(\ref{eq:f}). Whatever one wants to model with a left-right symmetric elementary one-dimensional PCA must be encoded in the choice of these parameters. We want to model single-species populations under logistic growth with such PCA. If we identify a cell in the state $0$ as an empty site or patch and a cell in the state $1$ as an individual or pack, the interpretation of the probabilities (\ref{eq:abcdef}) in terms of population dynamics is immediate. For example, $a$ represents the probability of spontaneous generation, $000 \to 010$, a somewhat unnatural biological process, while $f$ represents the probability of survival of the central individual ($111 \to 111$) in an overcrowded neighborhood. Not every set of values for the parameters yield sensible models from a population dynamics point of view, though. Contextual interpretations of these parameters are discussed in Sec.~\ref{sec:examples}.

\begin{table}
\caption{\label{tab:abcdef}Rule table for the general left-right symmetric one-dimensional elementary PCA. The first row lists the initial neighborhood and the other two rows give the probability at which the central cell reaches the state given in the leftmost column; cf.~(\ref{eq:a})--(\ref{eq:f}).}
\begin{ruledtabular}
\begin{tabular}{ccccccccc}
${}$ & $111$ & $110$ & $101$ & $100$ & $011$ & $010$ & $001$ & $000$ \\ \hline
$0$ & $1-f$ & $1-d$ & $1-e$ & $1-b$ & $1-d$ & $1-c$ & $1-b$ & $1-a$ \\
$1$ &  $f$  &  $d$  &  $e$  &  $b$  &  $d$  &  $c$  &  $b$  &  $a$
\end{tabular}
\end{ruledtabular}
\end{table}

Equation (\ref{eq:p1}) for $P_{t}(x_{i}'=1)=P_{t}(1)$, in the PCA defined by the left-right symmetric transition probabilities (\ref{eq:abcdef}) reads
\begin{equation}
\begin{aligned}
\label{eq:pca}
P_{t+1}(1)
&= aP_{t}(0,0,0) + b\big[P_{t}(0,0,1) + P_{t}(1,0,0)\big] \\
&+ cP_{t}(0,1,0) + d\big[P_{t}(0,1,1)+P_{t}(1,1,0)\big] \\
&+ eP_{t}(1,0,1) + fP_{t}(1,1,1).
\end{aligned}
\end{equation}
Applying approximation (\ref{eq:mf}) to the probabilities on the right-hand side of (\ref{eq:pca}) with $P_{t}(1)=x_{t}$ and $P_{t}(0)=1-x_{t}$ we obtain
\begin{equation}
\label{eq:xxx}
\begin{aligned}
x_{t+1} =
a(1-x_{t})^{3}&+(2b+c)x_{t}(1-x_{t})^{2} \\
&+(2d+e)x_{t}^{2}(1-x_{t})+fx_{t}^{3}.
\end{aligned}
\end{equation}
In the next section we determine the conditions on the transition probabilities $a$, $b$, \ldots, $f$ that can turn (\ref{eq:xxx}) into one of the population growth models (\ref{eq:map}) or (\ref{eq:cubic}).


\section{\label{sec:pcalog}PCA models of logistic population growth}

\subsection{\label{sec:constraints}Constraints on the transition probabilities}

Although the set of transition probabilities (\ref{eq:abcdef}) are the most general possible for an elementary one-dimensional left-right symmetric PCA, when $\phi(1\,|\,000) = a \ne 0$ there is spontaneous generation of living organisms, an undesirable feature for biological population models. To make model (\ref{eq:xxx}) biologically sensible, thus, we must require that $a=0$. When we set $a=0$, Eq.~(\ref{eq:xxx}) becomes $x_{t+1} = x_{t}h(x_{t})$ with
\begin{equation}
\label{eq:hhh}
\begin{aligned}
h(x_{t}) = (2b+c)&+\big[(2d+e)-2(2b+c)\big]x_{t} \\
&+\big[(2b+c)-(2d+e)+f\big]x_{t}^{2}.
\end{aligned}
\end{equation}

Now note that the transition probabilities $b$ and $c$ always appear in (\ref{eq:xxx}) and (\ref{eq:hhh}) in the combination $2b+c$. This happens because in the mean-field approximation (\ref{eq:mf}) to (\ref{eq:pca}), $P_{t}(0,0,1)$, $P_{t}(0,1,0)$, and $P_{t}(1,0,0)$, which are respectively associated with the transition probabilities $\phi(1\,|\,001)=b$, $\phi(1\,|\,010)=c$, and $\phi(1\,|\,100)=b$, all give rise to the same term $x_{t}(1-x_{t})^{2}$. \textit{Ditto} for the transition probabilities $d$ and $e$, which always appear together like $2d+e$. It is thus convenient to lump these transition probabilities into new variables $u=2b+c$ and $v=2d+e$, both in the range $0 \leq u, v \leq 3$.

Comparing the mean-field Eq.~(\ref{eq:hhh}) for the dynamics of the PCA (\ref{eq:abcdef}) with models (\ref{eq:map}) and (\ref{eq:cubic}) for the growth of a single-species population, we see that the coefficients of the powers of $x_{t}$ in (\ref{eq:hhh}), which we will denote by $[\,1\,]$, $[\,x_{t}\,]$, and $[\,x_{t}^{2}\,]$, must obey one of the two following sets of constraints:

\paragraph*{\textbf{Case I (Logistic map).}}
To recover (\ref{eq:map}) from (\ref{eq:hhh}) we must have $[\,x_{t}^{2}\,] = 0$ together with $[\,1\,] > 0$ and $[\,x_{t}\,] < 0$. These constraints translate into the following conditions on the transition probabilities of the PCA:
\begin{subequations}
\label{eq:noallee}
\begin{align}
\label{eq:nou}
u &> 0, \\
\label{eq:nouv}
-2u+v &< 0, \\
\label{eq:nouvf}
u-v+f &= 0.
\end{align}
\end{subequations}

\paragraph*{\textbf{Case II (Cubic map with weak Allee effect).}}
In this case, to recover the $g^{(+)}(x_{t})$ model in (\ref{eq:cubic}) from (\ref{eq:hhh}), $[\,x_{t}^{2}\,]$ must be negative while $[\,1\,]$ and $[\,x_{t}\,]$ must be positive. In terms of the transition probabilities these conditions read
\begin{subequations}
\label{eq:withallee}
\begin{align}
\label{eq:withu}
u &> 0, \\
\label{eq:withuv}
-2u+v &> 0, \\
\label{eq:withuvf}
u-v+f &< 0.
\end{align}
\end{subequations}

Note that because $2b+c \geq 0$ in (\ref{eq:hhh}) must be compared with the independent term of $g^{(\pm)}(x_{t})$ in (\ref{eq:cubic}), the PCA is not able to reproduce $g^{(+)}(x_{t})$, for which $g^{(+)}(0)=-r<0$, at least not in the single-cell mean-field approximation. Otherwise, if $2b+c=0$, i.\,e., if $b=c=0$, $h(x_{t})$ in (\ref{eq:hhh}) becomes
\begin{equation}
\label{eq:limit}
h(x_{t})=(2d+e)x_{t}(1-x_{t}).
\end{equation}
This form for $h(x_{t})$ corresponds to the limit $A \to 0$ in (\ref{eq:cubic}). Indeed, when $x_{t}/A \gg 1$, we can take $x_{t}/A \pm 1 \approx x_{t}/A$ and obtain $g^{(\pm)}(x_{t}) \approx (r/A)x_{t}(1-x_{t}/K)$, as in (\ref{eq:limit}). Note that as $A \to 0$, $r \to 0$ such that $r/A \to \mathrm{const}$. In this paper we gloss over this limiting case and always take $A>0$.


\subsection{\label{sec:examples}Analysis of the constraints and some examples}

One can argue that from the point of view of population dynamics the more interesting PCA models are those with small values of $e$ ($101 \to 111$) and $f$ ($111 \to 111$), to model the decrease of the birth rates in overcrowded neighborhoods (the logistic effect), and moderate to large values of $b$ ($001 \to 011$ and $100 \to 110$) and $d$ ($011 \to 011$ and $110 \to 110$), to model reproduction and survival under favorable conditions. The magnitude of $c$ ($010 \to 010$) depends on whether one wants to model individuals more or less resilient to loneliness and its consequences. Possible choices for the transition probabilities encompassing these arguments would be, for example, $e=f=b/2$ and $c=d$ or $d/2$, among others. In this subsection we analyze the two sets of constraints identified in Sec.~\ref{sec:constraints} and give examples of PCA that fall into each case. As we shall see, the classification of PCA into Case~I or II makes sense, since the analysis of their transition probabilities can be interpreted as describing the dynamics of a population under logistic growth {with or without} Allee effect.

\begin{figure}
\centering
\includegraphics[viewport=10 0 480 460, scale=0.35, clip]{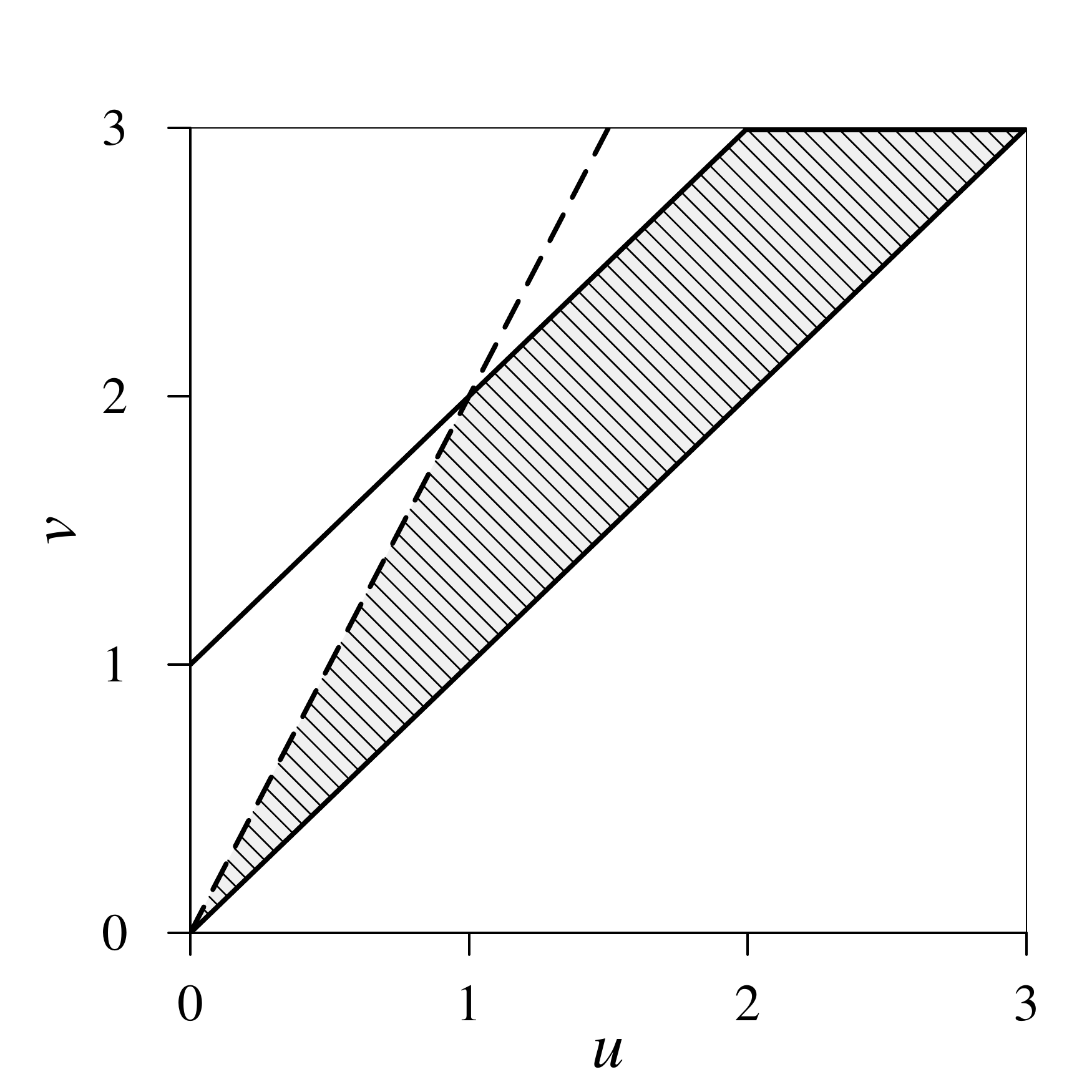}
\caption{\label{fig:si}Projection on the $uv$-plane of the solution set $S_{\mathrm{I}}$ of (\ref{eq:si}) (hatched area). Given $(u,v) \in S_{\mathrm{I}}$, $f$ is given by the plane $f=v-u$ ``hovering'' over the hatched area. The hatched polygon has vertices at $(u,v)=(0,0)$, $(1,2)$, $(2,3)$, and $(3,3)$, with total area $2/9$. The dashed edge $(0,0)$--$(1,2)$ as well as the lines external to the hatched area do not belong to $S_{\mathrm{I}}$.}
\end{figure}

\begin{figure}
\centering
\includegraphics[viewport=80 80 460 440, scale=0.535, clip]{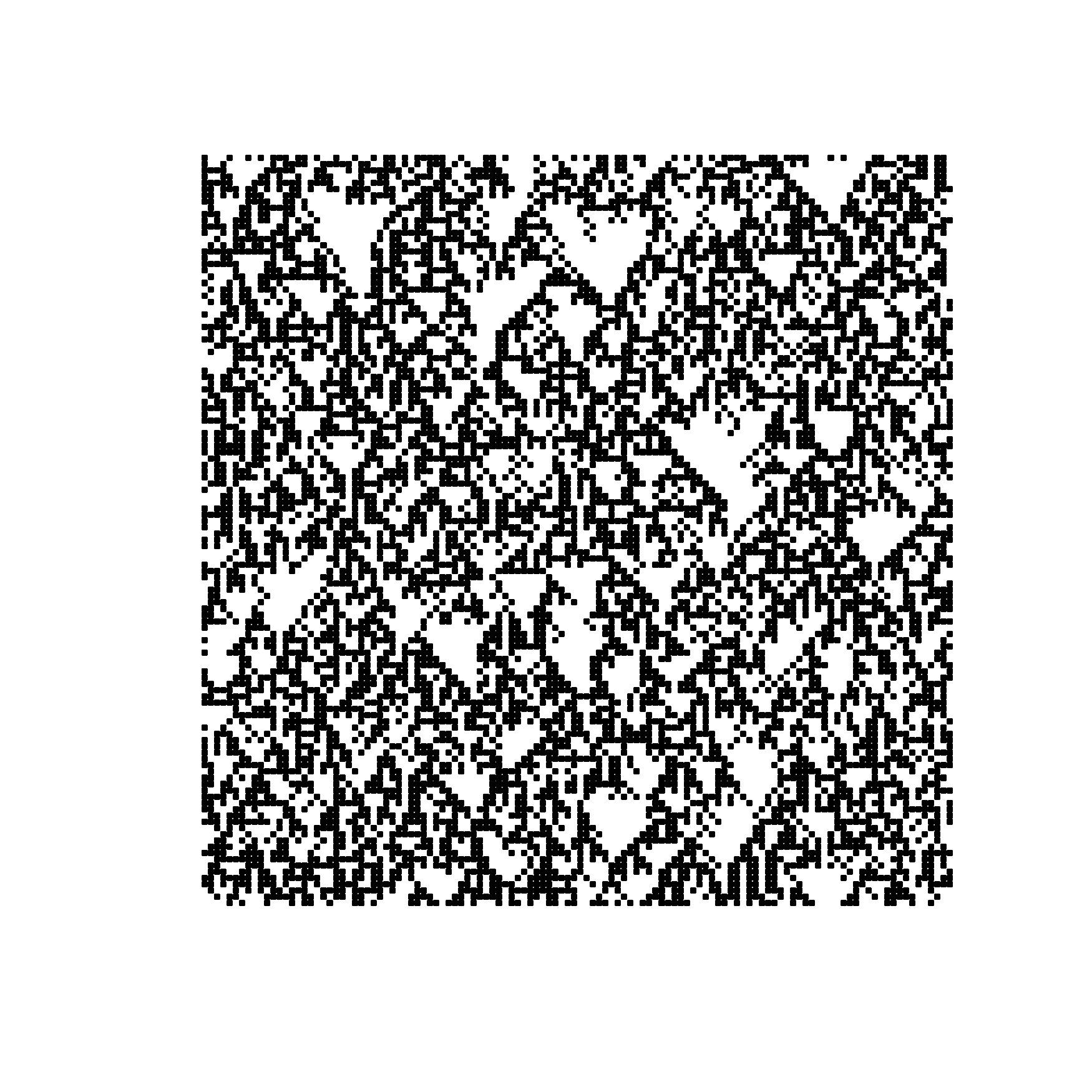}
\caption{\label{fig:pcai}Space-time diagram of PCA $(a,b,c,d,e,f)=(0,\frac{2}{3},\frac{1}{3},\frac{7}{8},\frac{1}{4},\frac{1}{3})$ exemplifying Case~I. A total of $120$ cells under periodic boundary conditions are evolved for $120$ time steps~(time runs downward) from an initially random state of density $1/2$.}
\end{figure}

\paragraph*{\textbf{Case I (Logistic map).}}
Constraints (\ref{eq:noallee}) can be recast as
\begin{equation}
\label{eq:si}
v < 2u, \quad 0 < u \leq v \leq 1+u,
\end{equation}
with $f=v-u$ always in the range $[0,1]$. Problem (\ref{eq:si}) is effectively a two-dimensional problem. Figure~\ref{fig:si} displays the projection $\{v<2u\}\cap\{0 < u \leq v \leq 1+u\}$ of the solution set $S_{\mathrm{I}}$ on the $uv$-plane. The solution set $S_{\mathrm{I}}$ includes the plane $f=v-u$, which, however, is not displayed in Fig.~\ref{fig:si}. The $uv$-projection of $S_{\mathrm{I}}$ occupies a considerable fraction (namely, $2/9$) of the $[0,3] \times [0,3]$ box available. That means that it is not hard to find a set of transition probabilities satisfying (\ref{eq:si}) that also makes sense in terms of population dynamics. For example, we can take $b=2/3$ ($001 \to 011$ and $100 \to 110$) to represent relatively high, but not certain, reproduction in a relatively favorable neighborhood, $c=1/3$ ($010 \to 010$) and $d=7/8$ ($011 \to 011$ and $110 \to 110$) to represent survival, with $c$ smaller than $d$ because of the ``loneliness factor,'' and $e=1/4$ ($101 \to 111$) again to represent possible but not certain (in fact, unlikely) reproduction into a crowded neighborhood. We then obtain $u=2b+c=5/3$, $v=2d+e=2$, and $f=v-u=1/3$ ($111 \to 111$) representing the possibility of survival in what we may call an overcrowded neighborhood. A space-time diagram of the resulting PCA is displayed in Fig.~\ref{fig:pcai} for a PCA of length $L=120$ initially occupied by $L/2=60$ individuals randomly distributed among the cells.

The equivalent logistic map (\ref{eq:map}) for the above choice of transition probabilities has $r=u=5/3$, within the interval of stability \cite{may1974,hassell,may1976a,may1976b,iterated,devaney}, and $K=u/(2u-v)=5/4$, namely,
\begin{equation}
\label{eq:maprK}
x_{t+1}=\frac{5}{3}x_{t}\Big(1-\frac{4}{5}x_{t}\Big).
\end{equation}
The stationary density of the logistic map (\ref{eq:maprK}) is given by $x_{\infty}=1/2$, while the average stationary density of the PCA measured from small scale simulations is $\langle x_{\infty}^{(\mathrm{PCA})} \rangle \simeq 0.484$.

\begin{figure}[t]
\centering
\includegraphics[viewport=100 130 460 380, scale=0.55, clip]{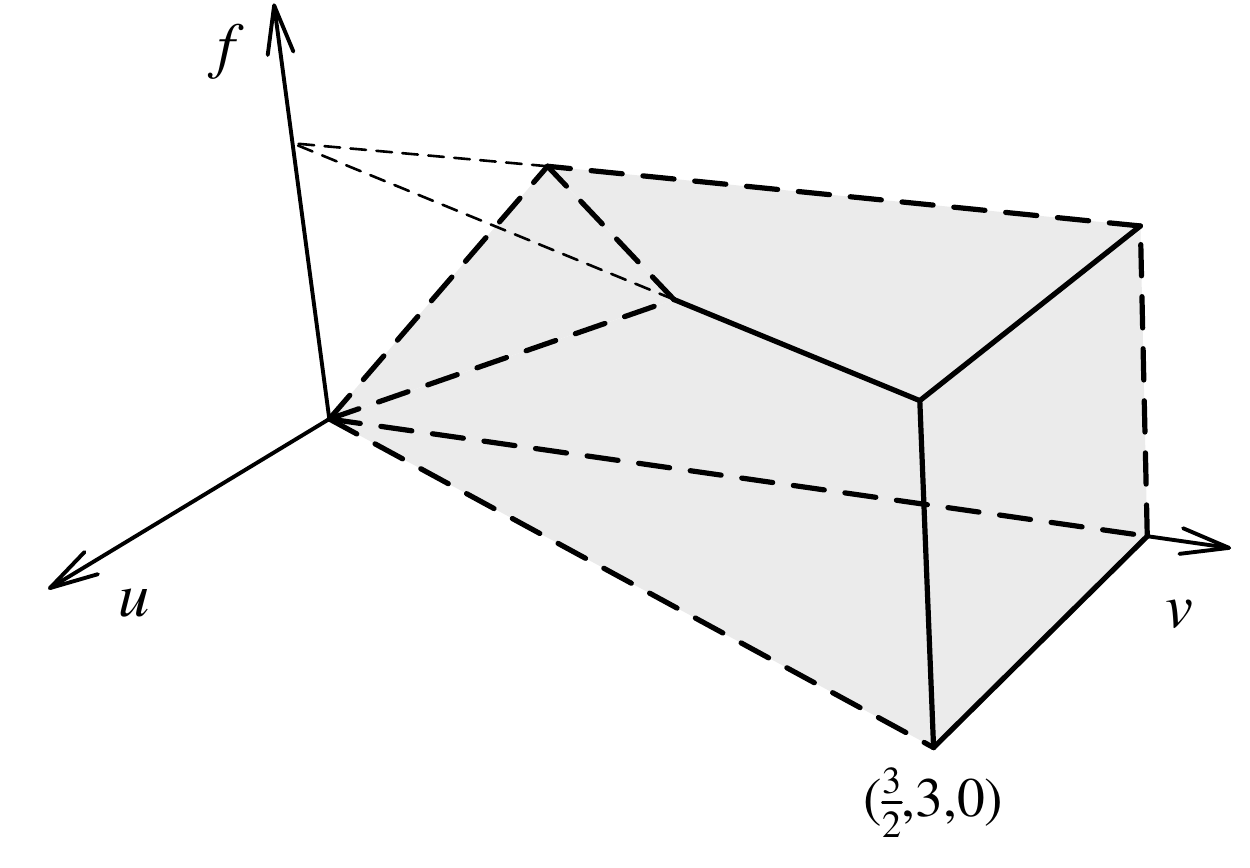}
\caption{\label{fig:sii}Solution set $S_{\mathrm{II}}$ (shaded volume) of (\ref{eq:sii}). The simplex has vertices at $(u,v,f)=(0,0,0)$, $(\frac{3}{2},3,0)$, $(0,3,0)$ (lower face), $(1,2,1)$, $(\frac{3}{2},3,1)$, $(0,3,1)$, $(0,1,1)$ (upper face), with total volume $\abs{S_{\mathrm{II}}}=25/12$. Dashed edges as well as faces bounded by dashed edges do not belong to $S_{\mathrm{II}}$.}
\end{figure}

\paragraph*{\textbf{Case II (Cubic map with weak Allee effect).}}
Constraints (\ref{eq:withallee}) are equivalent to
\begin{equation}
\label{eq:sii}
0 < 2u < v, \quad f < v-u.
\end{equation}
The solution set $S_{\mathrm{II}}$ of (\ref{eq:sii}) can be constructed as follows. In a coordinate system $(u,v,f)$, the first constraint $0<2u<v$ determines a triangular slab delimited by the planes $u=0$, $v=3$ and $v=2u$ contained in the box $(0,3] \times (0,3] \times [0,1]$. The second constraint defines a half-space bounded above by the plane $f=v-u$, i.\,e., by a plane normal to the direction $(1,-1,1)$. The intersection of this half-space with the triangular slab is $S_{\mathrm{II}}$. This set is depicted in Fig.~\ref{fig:sii}.

\begin{figure}[t]
\centering
\includegraphics[viewport=80 80 460 440, scale=0.535, clip]{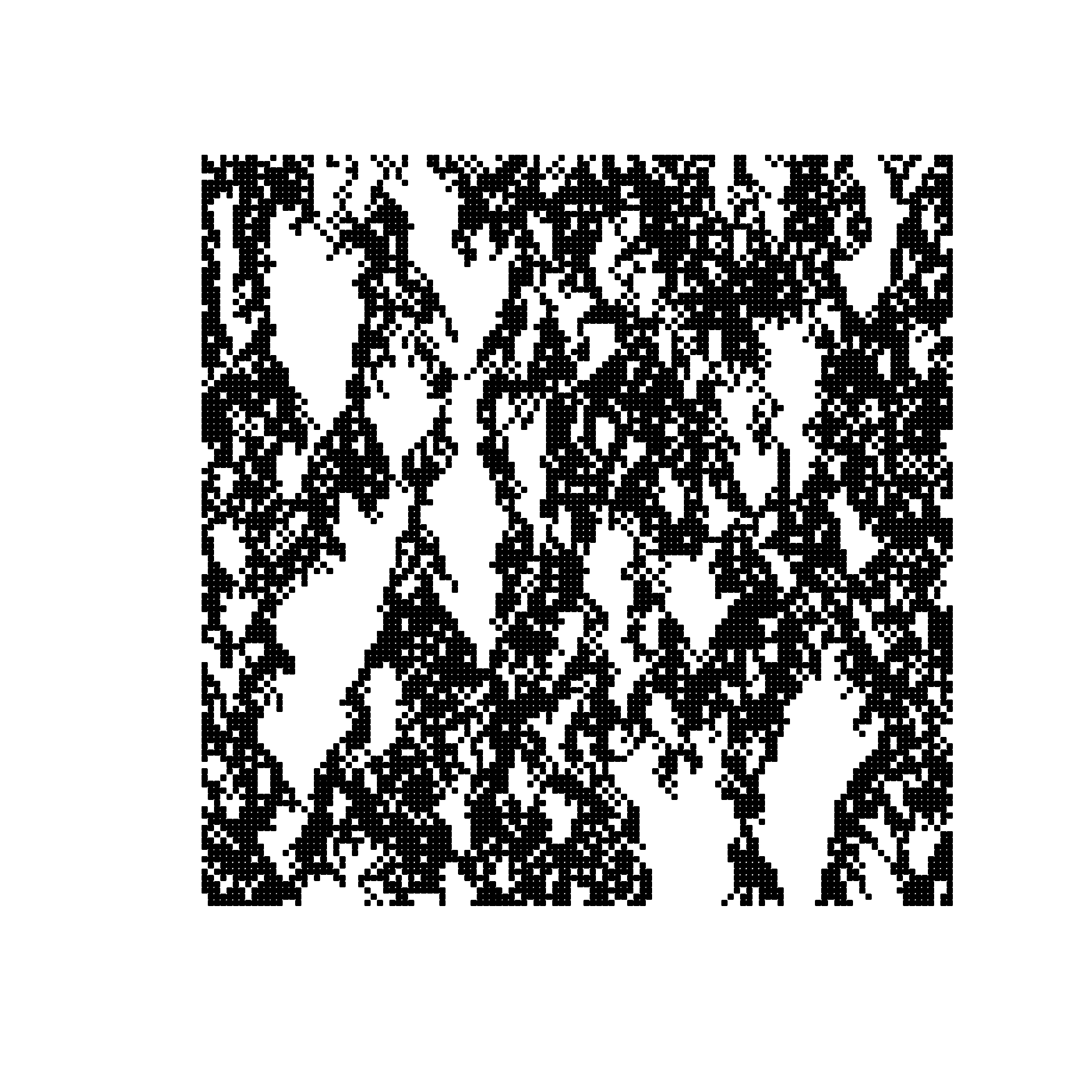}
\caption{\label{fig:pcaii}Space-time diagram of PCA $(a,b,c,d,e,f)=(0,\frac{3}{8},\frac{2}{5},\frac{4}{5},\frac{3}{4},\frac{4}{5})$ exemplifying Case~II. A total of $120$ cells under periodic boundary conditions are evolved for $120$ time steps~(time runs downward) from an initially random state of density $1/2$.}
\end{figure}

The volume of $S_{\mathrm{II}}$ is $25/12$, approximately $23\%$ of the total volume available, so that there are plenty of transition probabilities to choose from. In this case, however, choices with large values of $b$ ($001 \to 011$ and $100 \to 110$), which we argued formerly represent populations better prepared to thrive, are less numerous, since $u < 3/2$ in $S_{\mathrm{II}}$, entailing $b<3/4$, that moreover can be large only at the expense of $c$ ($010 \to 010$). In other words, for a given valid $u$, if we pick $b>1/2$ we necessarily have to pick $c<1/2$ and vice-versa; we cannot have both probabilities above $1/2$ as in Case~I. For instance, if we pick $b=2/3$, as in the previous example, we have to pick $c<1/6$, meaning that, on average, less than one out of six lone individuals ($010$) survive to tell the history. This is a manifestation of the Allee effect, seen from the point of view of the local microscopic processes. Another manifestation of the Allee effect in the opposite direction is reflected on the possible choices for $v$, which can be as large as the maximum possible, namely, $v=3$. Large values of $v$ means large values of $d$ ($011 \to 011$ and $110 \to 110$) and $e$ ($101 \to 111$), which can be interpreted as improved survival probability due to collective support despite the logistic (overcrowding) effect. Whatever the value of $v$, the transition probabilities $b$ ($001 \to 011$ and $100 \to 110$) and $c$ ($010 \to 010$) are bound by the condition $u<v/2$, further limiting the chances of ``unassisted'' reproduction ($b$) and survival of lone individuals ($c$). Figure~\ref{fig:pcaii} depicts the space-time diagram of a Case~II PCA of length $L=120$ with parameters $(a,b,c,d,e,f)=(0,\frac{3}{8},\frac{2}{5},\frac{4}{5},\frac{3}{4},\frac{4}{5})$ initially occupied by $L/2=60$ individuals at random cells.

Note how the clusters of individuals in Fig.~\ref{fig:pcaii} are more compact than the clusters in Fig.~\ref{fig:pcai}. Also note how some of the clusters are fringed by checkerboard-like regions of low populational density before the plains devoid of individuals. This kind of pattern is typical of invasion processes, in which Allee effects play a major role \cite{cellscale,mcelwain,kareiva,hastings,gruntfest}. Invasion processes, however, are better modeled in two spatial dimensions.

The pattern displayed by the PCA in Fig.~\ref{fig:pcaii} strongly resembles the pattern of directed (site or bond) percolation clusters at or slighly above criticality \cite{durrett,dickman,haye}. The relationship of these PCA (both Case~I and II) with percolation processes as well as with other known CA and PCA, in particular with the Domany-Kinzel (DK) PCA \cite{dkpca}, is an interesting question. For example, the map to the DK PCA would require that $c=0$; exact correspondence would additionally require that $b=d$ and $e=f$. A PCA with the same parameters as the one in Fig.~\ref{fig:pcaii} but with $c=0$ would roughly correspond to an inactive instance of the DK PCA. It thus seems that sustained activity of the PCA in Case~II depends to a certain extend on the survival probability of lone individuals, as embodied by the transition probability $c=\phi(1\,|\,010)$. Preliminary analysis suggests that the dynamics of the PCA is not very sensitive to the particular values of $c$ as long as $c>0$.

\section{\label{sec:pq}Single-parameter PCA}

Most PCA models in the literature are one and two-parameter models, since they are already rich and flexible enough to model a great many phenomena, in one or more dimensions, and display a variety of nontrivial behavior, from disordered and chaotic phases to phase transitions of several different kinds \cite{wolfram,toffoli,chopard,boccara,adamatzky,dickman,haye,arXiv,nazim}. It is thus not entirely without interest to identify single-parameter PCA that in the single-cell mean-field approximation yield models for the logistic growth of populations, with or without weak Allee effects.


\subsection{\label{sec:param}Parametrization of the transition probabilities}

We can reduce the number of parameters of PCA (\ref{eq:abcdef}) from six to a single one, say, $p \in \mathbb{R}$, by parametrizing the transition probabilities like $a(p)$, $b(p)$, \ldots, $f(p): \mathbb{R} \to [0,1]$. The simplest possible parametrization is by linear functions, $a(p)=a_{0}+a_{1}p$, $b(p)=b_{0}+b_{1}p$, \ldots, $f(p)=f_{0}+f_{1}p$, with $p \in [p_{1},p_{2}]$. If we choose $p \in [0,1]$, the transition probabilities become then given either by $p$ or $q=1-p$ and there are $2^{6}=64$ possible parametrizations for the transition probabilities, namely, $(a,b,c,d,e,f)=(p,p,p,p,p,p)$, $(p,p,p,p,p,q)$, \ldots, $(q,q,q,q,q,q)$. Now, note that since $q=1-p \Leftrightarrow p=1-q$, if a given parametrization has solution set $p \in (p_{1},p_{2}]$ (we take a half-open, half-closed interval for the sake of illustration), interchanging the roles of $p$ and $q$ in the parametrization leads to the ``dual'' parametrization with solution set $q \in (p_{1},p_{2}]$, that is, $p \in [1-p_{2},1-p_{1})$. As a consequence, as $p$ ranges over $[1-p_{2},1-p_{1})$, the transition probabilities of the dual parametrization range over the same values as the initial parametrization when $p$ ranges over $(p_{1},p_{2}]$, and the two parametrizations are equal. We then have, in fact, a maximum of $32$ unique parametrizations.

In our study, $a(p)=0$ and we end up with only $16$ possible parametrizations of the transition probabilities. Again, it is convenient to look at these parametrizations through the variables $u=2b+c$ and $v=2d+e$. We now determine for which values of $p$ PCA (\ref{eq:abcdef}) comply with constraints (\ref{eq:noallee}) and (\ref{eq:withallee}).

\paragraph*{\textbf{Case I (Logistic map).}}
In this case $f=(2d+e)-(2b+c)=f(b(p),\,c(p),\,d(p),\,e(p))$, and we have to examine only the eight possible parametrizations for $b$, $c$, $d$, and $e$. We then enforce constraints (\ref{eq:noallee}) on the parametrized transition probabilities and obtain the valid ranges for $p$ in each case. The solution sets $S_{\mathrm{I}}(a,b,c,d,e,f;p)$ found are summarized in Table~\ref{tab:si}. Let us look at one example closely to make Table~\ref{tab:si} more clear. If we take $(a,b,c,d)=(p,p,q,p)$, we have $u=2b+c=3p$ and $v=2d+e=2-p$. The constraint $u>0$ (\ref{eq:nou}) implies $p>0$, while the constraint $2u>v$ (\ref{eq:nouv}) implies $p>2/7$. Since $f=v-u=2-4p$ must be a number between $0$ and $1$, we further require that $1/4 \leq p \leq 1/2$. The final solution set is the interval $p \in (2/7,1/2]$. Solutions can also be read as line segments in $S_{\mathrm{I}}$. In our example, the equation for the line segment in symmetric form is $u/3=2-v=(2-f)/4=p$, with $p \in (2/7,1/2]$.

\begin{table}
\caption{\label{tab:si}Solution sets for the one-parameter version of PCA in Case~I (logistic map) according to the possible parametrizations indicated in the first column. The solution sets are given as line segments (equations in symmetic form) in the $(u,v,f)$ space. In every case $f=(2d+e)-(2b+c)=v-u$.}
\begin{ruledtabular}
\begin{tabular}{cc}
$(a,b,c,d,e,f)$ & $S_{\mathrm{I}}(a,b,c,d,e,f;p)$ \\
\hline
$(0,p,p,p,p,*)$ & $u/3=v/3 \in (0,1]$, $f=0$ \\
$(0,p,p,p,q,*)$ & $u/3=v-1=(1-f)/2 \in (\frac{1}{5},\frac{1}{2}]$ \\
$(0,p,p,q,p,*)$ & $u/3=2-v=(2-f)/4 \in (\frac{2}{7},\frac{1}{2}]$ \\
$(0,p,p,q,q,*)$ & $u/3=(3-v)/3=(3-f)/6 \in (\frac{1}{3},\frac{1}{2}]$ \\
$(0,p,q,p,p,*)$ & $u-1=v/3=(1+f)/2 \in [\frac{1}{2},1]$ \\
$(0,p,q,p,q,*)$ & $u-1=v-1 \in[0,1]$, $f=0$ \\
$(0,p,q,q,p,*)$ & $u-1=2-v=(1-f)/2 \in (0,\frac{1}{2}]$ \\
$(0,p,q,q,q,*)$ & $u-1=(3-v)/3=(2-f)/4 \in [\frac{1}{4},\frac{1}{2}]$
\end{tabular}
\end{ruledtabular}
\end{table}

\paragraph*{\textbf{Case II (Cubic map with weak Allee effect).}}

Here we have to scrutinize all $16$ different parametrizations of the transition probabilities $b$, $c$, $d$, $e$ and $f$ to get the whole picture. However, from constraint (\ref{eq:withuvf}) we see that all instances with $u=v$ are forbidden, ruling out the four parametrizations $(0,p,p,p,p,*)$ and $(0,p,q,p,q,*)$. Four other parametrizations violate constraint (\ref{eq:withuv}), namely, $(0,p,q,p,p,*)$ and $(0,p,q,q,p,*)$, because the first ones require that $p>2$ and the second ones that $p<0$. Upon examination, we found that parametrization $(0,p,p,p,q,q)$ is also impossible because it would also require that $p<0$. The resulting solution sets are listed in Table~\ref{tab:sii}.


\subsection{\label{sec:mixed}Mixed CA and PCA}

\begin{table}[t]
\caption{\label{tab:sii}Solution sets for the one-parameter version of PCA in Case~II (cubic map with weak Allee effect) according to the possible parametrizations indicated in the first column. The solution sets are given as line segments (equations in symmetic form) in the $(u,v,f)$ space. The last column gives the mixed rule specification of the PCA.}
\begin{ruledtabular}
\begin{tabular}{ccc}
$(a,b,c,d,e,f)$ & $S_{\mathrm{II}}(a,b,c,d,e,f;p)$ & Mixed PCA rule \\ \hline
$(0,p,p,p,q,p)$ & $u/3=v-1=f \in (0,\frac{1}{5})$ & $p222$--$q32$ \\
$(0,p,p,q,p,p)$ & $u/3=2-v=f \in (0,\frac{2}{7})$ & $p182$--$q72$ \\
$(0,p,p,q,p,q)$ & $u/3=2-v=1-f \in (0,\frac{2}{7})$ & $p54$--$q200$ \\
$(0,p,p,q,q,p)$ & $u/3=(3-v)/3=f \in (0,\frac{1}{3})$ & $p150$--$q104$ \\
$(0,p,p,q,q,q)$ & $u/3=(3-v)/3=1-f \in (0,\frac{1}{3})$ & $p22$--$q232$ \\
$(0,p,q,q,q,p)$ & $u-1=(3-v)/3=f \in [0,\frac{1}{5})$ & $p146$--$q108$ \\
$(0,p,q,q,q,q)$ & $u-1=(3-v)/3=1-f \in [0,\frac{1}{5})$ & $p18$--$q236$
\end{tabular}
\end{ruledtabular}
\end{table}

The one-parameter PCA in Case~II are examples of mixed PCA that have appeared in the literature before \cite{nazim,p182q200,pxor}. In a mixed PCA, two or more deterministic CA rules are combined probabilistically such that sometimes one rule is applied, some other times another rule is applied to a given cell. Mixed PCA are closely related with asynchronous PCA \cite{nazim}: in an asynchronous PCA, one of the rules is the identity map $x_{i}^{t+1}=x_{i}^{t}$. We can specify mixed PCA by telling which rules are applied with which probabilities. For example, the rule table for PCA $(a,b,c,d,e,f)=(0,p,p,q,p,q)$, third line in Table~\ref{tab:sii}, appears in Table~\ref{tab:ppqpq}. Reading the lines of the table as binary numbers (Wolfram's encoding \cite{wolfram}) we obtain that PCA $(0,p,p,q,p,q)$ is the mixed PCA $p54$--$q200$. In \cite{p182q200}, the authors studied the closely related mixed PCA $p182$--$q200$. The two models differ in the probability for the transition $111 \to 111$, that is $\phi(1\,|\,111)=p$ in PCA $p54$--$q200$ and $\phi(1\,|\,111)=1$ in PCA $p182$--$q200$. This single difference, however, implies that PCA $p182$--$q200$ has two absorbing configurations, the all-$0$ (empty lattice) and the all-$1$ (full lattice) configurations. PCA $p182$--$q200$ displays an extinction-survival phase transition at the critical point $p^{*} \simeq 0.488$ in the directed percolation universality class of critical behavior. PCA $p54$--$q200$ also has an extinction-survival phase transition between the empty lattice and a stationary state with finite positive density at $p^{*} \simeq 0.575$, out of the interval $p \in (0,\frac{1}{7})$ in which its mean-field approximation yields the logistic growth model with weak Allee effect. Estimates of the critical points for the mixed PCA of Table~\ref{tab:sii} are given in the Appendix.

\begin{table}
\caption{\label{tab:ppqpq}Rule table for PCA $(a,b,c,d,e,f)=(0,p,p,q,p,q)$, third line in Table~\ref{tab:sii}. The first row lists the initial neighborhood and the other two rows give the state that the central cell reaches with the probability given in the first column. This is mixed PCA $p54$--$q200$.}
\begin{ruledtabular}
\begin{tabular}{ccccccccc}
${}$ & $111$ & $110$ & $101$ & $100$ & $011$ & $010$ & $001$ & $000$ \\ \hline
$p$ &  $0$  &  $0$  &  $1$  &  $1$  &  $0$  &  $1$  &  $1$  &  $0$  \\
$q$ &  $1$  &  $1$  &  $0$  &  $0$  &  $1$  &  $0$  &  $0$  &  $0$
\end{tabular}
\end{ruledtabular}
\end{table}


\section{\label{sec:summary}Summary and conclusions}

We explored the modeling possibilities of the most general right-left symmetric one-dimensional elementary PCA to the dynamics of a single-species unstructured population with nonoverlapping generations. Our results consist in the classification of all sets of parameters of the PCA that furnishes in first-order mean-field approximation either the logistic map for the density of a population or a cubic map that describes the dynamics of a population under weak Allee effects. The PCA that give rise to the logistic map, for example, are composed of microscopic transitions that describe individuals that struggle against overcrowded neighborhoods, that hamper their chances of reproducing and survival (small probabilities for transitions like $101 \to 111$ and $111 \to 111$) and prefer more capacious environments (large probabilities for transitions like $010 \to 010$ and $100 \to 110$). In the same manner, the PCA that furnish the cubic map for the dynamics of a population under weak Allee effects is made of microscopic transitions that describe individuals that prefer to team up as long as the neighborhood is not too crowded (relatively high probabilities for events like $011 \to 011$ and $101 \to 111$) but suffer from loneliness (smaller chances that $001 \to 011$ and $010 \to 010$). Examples were given in Sec.~\ref{sec:examples}.

In Sec.~\ref{sec:pq} we obtained all one-parameter PCA that yield the logistic map for the dynamics of a population density $x_{t}$ in the mean-field approximation, both in the absence (Case~I) and in the presence (Case~II) of weak Allee effects. We found eight different PCA in the first case and seven in the second case. The PCA in Case~II can be viewed as probabilistic mixtures of two different CA rules. All of them display phase transitions between an inactive ($x_{\infty}=0$) and an active ($x_{\infty}>0$) phase, however at critical points out of the range in which their single-cell mean-field approximation becomes the logistic growth with weak Allee effects. It would be interesting to investigate the phase transitions in these models irrespective of their connection with the logistic or cubic maps of this work, as well as their relationship with percolation processes and other known CA and PCA.

The analysis of the discrete maps found in this paper was left aside for two reasons. First, because the logistic map (\ref{eq:map}) is one of the most well studied dynamical systems ever, and we do not feel compelled to add anything to the extensive body of literature concerning its behavior \cite{may1974,hassell,may1976a,may1976b,iterated,devaney}. Second, because the dynamics of the full cubic map (\ref{eq:xxx}) and its interpretation in the context of population dynamics and Allee effects deserve a detailed investigation that we intend to carry out elsewhere. Note that although cubic maps have appeared in mathematical biology before \cite{annny,cubic,npbss}, in this paper they are obtained from microscopic models of population dynamics involving the behavior of individual agents. One should not expect that the bifurcations displayed by the logistic and cubic maps will appear in a PCA with the same mean-field equation. Mean-field models, as well as the maps, are ``well stirred,'' while the PCA dynamics is local. One way to recover the bifurcations is to add some reshuffling (long-range mixing) among the interacting particles or to rewire a fraction of the cells to explore the small-world effect (see \cite{pppe,arXiv,popcycles} for examples, discussion, and references).

Clearly, ecological complexity, even at the limited scale of a single species within a single patch, cannot be grasped by a one-dimensional PCA. The extension of the single-cell mean-field approximation to two-dimensional models, with two or more states per cell, together with automated search for meaningful combinations of parameters may prove fruitful.


\begin{acknowledgments}

We thank library specialist Angela K.\ Gruendl (UIUC) for providing a copy of Ref.~\cite{dobrushin} and one anonymous reviewer for constructive
comments improving the manuscript. J.\,R.\,G.\,M.\ acknowledges the S\~{a}o Paulo State Research Foundation, FAPESP (Brazil), for partial support under grant 2015/21580-0; Y.\,G.\ acknowledges the Coordena\c{c}\~{a}o de Aper\-fei\-\c{c}oa\-mento de Pessoal de N\'{\i}vel Superior, CAPES (Brazil), for support from a Ph.\,D. scholarship.

\end{acknowledgments}


\appendix*

\section{\label{app:a}Phase transitions in the mixed PCA of Table~\ref{tab:sii}}

Here we obtain the critical points of the extinction-survival phase transition of the PCA listed in Table~\ref{tab:sii} from their mean-field Eq.~(\ref{eq:xxx}) and small scale numerical simulations.

\begin{table}[b]
\caption{\label{tab:pc}Critical points for the single-parameter mixed PCA listed in Table~\ref{tab:sii} at the single-cell mean-field approximation ($p^{*}_{\mathrm{mf}}$) and from small-scale numerical simulations ($p^{*}_{\mathrm{sim}}$). Uncertainties in the values of $p^{*}_{\mathrm{sim}}$ are approximately $\pm 0.001$.}
\begin{ruledtabular}
\begin{tabular}{ccc}
Mixed PCA rule & $p^{*}_{\mathrm{mf}}$ & $p^{*}_{\mathrm{sim}}$ \\ \hline
$p222$--$q32$  & $1/3$ & $0.543$ \\
$p182$--$q72$  & $1/3$ & $0.563$ \\
$p54$--$q200$  & $1/3$ & $0.576$ \\
$p150$--$q104$ & $1/3$ & $0.573$ \\
$p22$--$q232$  & $1/3$ & $0.589$ \\
$p146$--$q108$ &  $0$  & $0.648$ \\
$p18$--$q236$  &  $0$  & $0.676$
\end{tabular}
\end{ruledtabular}
\end{table}

The critical point in the mean-field approximation can be calculated as follows. In the stationary state $x_{t+1}=x_{t}=x_{\infty}$, and from (\ref{eq:xxx}) with $a=0$ we obtain that either $x_{\infty}=0$ or
\begin{equation}
\label{eq:quadratic}
(u-v+f)x^{2}_{\infty}+(v-2u)x_{\infty}+(u-1)=0,
\end{equation}
where $u$, $v$, and $f$ are the variables introduced in Sec.~\ref{sec:constraints}. Equation (\ref{eq:quadratic}) has solutions
\begin{equation}
\label{eq:xpm}
x^{(\pm)}_{\infty}=\frac{-(v-2u) \pm \sqrt{(v-2u)^{2}-4(u-v+f)(u-1)}}{2(u-v+f)}.
\end{equation}
Now we must find the critical points $p^{*}$ at which $x^{(\pm)}_{\infty}(p^{*})=0$. The candidate roots of (\ref{eq:xpm}) are easily seen to be the points at which $4(u-v+f)(u-1)=0$; however, since $u-v+f$ is a removable singularity of (\ref{eq:xpm}), the candidate roots are actually obtained from the condition $u-1=0$. The critical point $p^{*}_{\mathrm{mf}}$ is then given by the root between $0$ and $1$ and the mean-field stationary density profile is the solution of (\ref{eq:xpm}) with $x_{\infty}(p) \geq 0$ for $p \geq p^{*}_{\mathrm{mf}}$ (or, sometimes, $p \leq p^{*}_{\mathrm{mf}}$). The mean-field critical points calculated for the mixed PCA of Table~\ref{tab:sii} are given in Table~\ref{tab:pc}, together with critical points $p^{*}_{\mathrm{sim}}$ estimated from simulations of PCA of $L=8000$ cells with data averaged over $4000$ samples of the stationary density.

With every continuous phase transition there comes the question of its universality class, i.\,e., of the critical exponents ruling its scaling behavior at criticality \cite{dickman,haye}. We guess, on the basis of the directed percolation conjecture \cite{janssen,grassberger}, that the critical behavior of the PCA of Table~\ref{tab:sii} all belong to the directed percolation universality class of critical behavior.


\end{document}